# Electronic and Magnetic Properties of Pristine and Hydrogenated Borophene Nanoribbons

Fanchen Meng,[1] Xiangnan Chen,[2] Songsong Sun,[3] and Jian He[1,*]

1. Department of Physics and Astronomy, Clemson University, Clemson, SC 29634-0978, USA

2. Transportation Equipment and Ocean Engineering College, Dalian Maritime University, Dalian, 116026, China

3. Key Laboratory of Advanced Technology of Materials (Ministry of Education), School of Physical Science and Technology, Southwest Jiaotong University, Chengdu 610031, China

\* Corresponding author. Email address: jianhe@g.clemson.edu

**Abstract**

The groundbreaking works in graphene and graphene nanoribbons (GNRs) over the past decade, and the recent discovery of borophene draw immediate attention to the underexplored borophene nanoribbons (BNRs). We herein report a density functional theory (DFT) study of the geometric, electronic and magnetic properties of BNRs as a function of orientation (denoted as BxNRs and ByNRs with the orientation along x- and y-axis, respectively), ribbon width ($N_x$, $N_y$ = 4 to 15), and hydrogenation effects. We found that the anisotropic quasi-planar geometric structure of BNR and its edge states profoundly govern the electronic and magnetic properties. Pristine ByNRs adopt a magnetic ground state, anti-ferromagnetic (AFM) or ferromagnetic (FM) depending on the ribbon width, while pristine BxNRs are non-magnetic (NM). Upon hydrogenation, all BNRs become NM. Interestingly, both pristine and hydrogenated ByNRs undergo a metal-semiconductor-metal transition at $N_y = 7$, while BxNRs remain metallic.

1. Introduction

The discovery of graphene [1] ushers in an era of two-dimensional (2-$d$) unit cell thick materials, ranging from boron nitride [2-5], transition -metal dichalcogenides (TMDCs) [6-9], phosphorene [10-12], to silicene [13-16]. These 2-$d$ materials exhibit exceptional electronic, magnetic, optical, and mechanical properties, promising innovative concepts and next-generation information technologies. In particular, the carbon-based atom-thick 2-$d$ and quasi-1-$d$ materials, namely, graphene and graphene nanoribbons (GNRs), have constituted a new paradigm in condensed matter physics and device engineering research [17-29]. From graphene to GNRs, the change is not merely morphological. Compared to graphene, GNRs exhibit many novel electronic and spin transport properties owing to the edge states [21, 27-29]. For example, GNRs with a width less than 10 nm possess an electronic band gap [21, 27, 28]. While graphene is a zero-gap semimetal, both zigzag graphene nanoribbons (ZGNRs) and armchair graphene nanoribbon (AGNRs) are semiconductors. [26] Furthermore, hydrogenated ZGNRs adopt an anti-ferromagnetic (AFM) ground state along with a ferromagnetic (FM) meta-stable state, hydrogenated AGNRs are found to be non-magnetic (NM), while graphene is NM. [26] With increasing ribbon width, GNRs crossovers to NM and the band gap diminishes.

Boron (B), carbon's nearest neighbor in the periodic table of chemical elements, exhibit rich crystal chemistry only next to carbon [30-42]. The bonding between B atoms is more complex than the case of carbon due to 2-center and 3-center B-B bonding configuration, giving rise to more than 16 allotropes. In particular, various allotropes of boron-based nanostructures have been predicted and investigated theoretically [32-39].

Experimentally, it is observed that B can form fullerene-like cage, $B_{40}^-$ [32]; the quasi- planar $B_{36}$ cluster with a central hexagonal vacancy [33]; icosahedral $B_{12}^-$ cage that is the building block of many bulk B structures [40]; highly aromatic $B_{13}^+$ [41]; and $B_{19}^-$, a nearly circular planar structure with a central $B_6$ pentagonal unit bonded to an outer $B_{13}$ ring [42]. Notably, the synthesis of unit-cell thin 2-*d* borophene, *the boron analogue of graphene*, was only successful until recently. Mannix *et al.* [43]

reported the synthesis of boronphene on Ag(111) substrate, a buckled triangular lattice without vacancies. Borophene adopts a corrugated quasi-planar structure, in contrast to the planar honeycomb hexagonal lattice of graphene [1] due to the electron deficiency of boron compared with carbon. The results of scanning tunneling microscope (STM) measurements further showed that borophene is a high anisotropic metal [43], in a vast contrast to the isotropic zero-gap semi-metallic graphene. In particular, the crystal structure derived by Mannix *et al.* immediately enabled a series of theoretical works. Peng *et al.* employed first principles calculations to study the electronic structure, bonding characteristics, as well as optical and thermodynamic properties of borophene [44]. They found that borophene has high optical transparency and electrical conductivity, making it a promising candidate for transparent conductors used in photovoltaics. Xu *et al.* [45] combined high-throughput screening with first principles calculations to demonstrate a novel growth mechanism of borophene from clusters, and ribbons on Ag(111). Penev *et al.* [46] proposed that the borophene formed on a metal substrate might exhibit intrinsic phonon-mediated superconductivity, with a critical temperature $T_c$ in the range 10–20 K. Zhang *et al.* [47] pointed out that the storage capacity of borophene based Li-ion and Na-ion batteries might be several times higher than the commercial graphite electrode and the highest among all the 2D materials discovered to date.

Briefly after Mannix's report, Feng *et al.* [48] reported the 2-*d* triangular lattice sheets $\beta_{12}$ and $\chi_3$ epitaxially grown on Ag (111) substrate without discernible vertical undulations. Notably, the crystal structure of $\beta_{12}$ and $\chi_3$ boron sheet is different from that identified by Mannix *et al.* [43].

In analogy with the research focus shift from graphene to GNRs mentioned above, it is thus compelling to study the electronic and magnetic properties of borophene nanoribbons (BNRs). BNRs might be realized either by cutting the synthesized borophene, or by patterning epitaxially grown borophenes. These methods have attained considerable success in preparing graphene nanoribbons (GNRs). [23-25]. Importantly, the study of BNRs is fundamentally interesting in light of the anisotropic crystal structure of borophene. The orientation of BNR thus becomes a control variable (tuning parameter). In addition and in general, the edge states are expected to play a profound role.

To date, the study of BNRs is sparse. Ding *et al.* [49] and Saxena *et al.* [50] early on investigated the electronic properties and the stability of boron nanoribbons using first principle calculations, however, the boron nanoribbons studied therein were based on hypothetical motifs rather than experimentally confirmed structures. Very recently, Zhang *et al.* [51] employed non-equilibrium Green's function (NEGF) combined with the first principles method to study the phonon transport properties of experimentally confirmed $\beta_{12}$ boron sheets and $\chi_3$ boron nanoribbons. Nonetheless, the study of electronic and magnetic properties of BNRs based on experimentally confirmed structure [43, 48] remains open. We will provide insight to this problem.

In this work, we conducted a systematical study on the geometric, electronic and magnetic properties of BNRs, based on the crystal structure derived by Mannix *et al.* [43]. The size effects, namely, the ribbon width effects on the geometric, electronic and magnetic properties, were carefully examined. Furthermore, we compared the pristine and hydrogenated BNRs to help elucidate the impact of edge states. This paper is organized as follows. Firstly, we introduce and validate the calculation methods applied in this work. The optimized crystal (geometric) structure of pristine BNRs is obtained. Secondly, the geometric, electronic and magnetic properties of pristine BNRs with different orientations and ribbon widths are discussed. In particular, we will compare the case of BNRs with GNRs on several occasions. Finally, we study the impact of edge states on the geometric, magnetic and electronic properties of BNRs by implementing hydrogenation.

## 2. Computational Methods

The equilibrium structures of BNRs are fully optimized by employing density functional theory (DFT) as implemented in the QUANTUM ESPRESSO package [52]. The generalized gradient approximation (GGA) [53] of Perdue-Burke-Ernzerhof (PBE) was used as the exchange-correlation functional. The interaction between electrons and ions is modeled by the ultrasoft pseudopotential. The cutoff for the kinetic energy was set to 50 Ry (1 Ry = 13.60569 eV) for the plane-wave expansion of the electronic wave functions. The charge-density cutoff was kept at 500 Ry. The Brillouin zone integration was performed using the Monkhorst-Pack scheme [54] with 13 meshes along periodic orientation. In order to eliminate the artificial image interaction, a vacuum layer of 20 Å thick is used along non-periodic direction. Spin-polarized calculations were employed for geometric optimization and physical properties calculations. All structures were fully optimized until the total energy converged to at least $10^{-6}$ Ry, and the inter-atomic forces became smaller than $10^{-4}$ Ry/bohr.

To validate the plane-wave method discussed above, parallel calculations were carried out using an ADF/BAND package [55, 56], which is also based on DFT. Unlike the plane wave method used in QUANTUM ESPRESSO, ADF/BAND adopts Slater-type local basis functions. In this calculation, Slater-type local basis functions of triple-ζ quality with one polarization function (TZP) were used as basis set and a small frozen core approach was adapted. The parallel spin-polarized calculations were also carried out using the PBE [53] exchange correlation functional within the generalized gradient approximation (GGA) for geometric optimization and physical properties calculations. The Wiesenekker-Baerends scheme [57] was used to sample k-point mesh over the Brillouin zone. The Becke fuzzy cells integration scheme [58] was employed, where the integration parameter set to "good", resulting in 59 irreducible k-points for the 2-*d* borophene. For the BNRs, the integration parameter was also set to "good" except for that the *k*-space sampling was kept fixed at 31 for consistency, resulting in total 16 k-points in the irreducible wedge. In order to take the scalar relativistic effects into account, Zero Order Regular Approximation (ZORA) [59] was included in all calculations.

## 3. Results and Discussions

**3.1 Geometric and electronic structure of borophene.** The geometric structures of BNRs are displayed in Fig. 1 (a)-(c). The optimized lattice parameters of borophene are **a** = 1.616 Å and **b** = 2.868 Å with space group Pmmn (tetragonal), in comparison, the space group of graphene is P6/mmm (hexagonal). The calculated thickness is ~ 0.9 Å (Fig. 2 (b) and (c)). It should be noted that (*i*) unlike graphene that is only one-atom thick, borophene adopts a quasi-planar corrugated structure; and (*ii*) unlike graphite and graphene, bulk boron can't be attained by simple stacking borophene. The calculated electronic band structure indicates that borophene is a metallic along Γ-Y direction but has a band gap along Γ-X direction (Fig. 2 (d)). Both the calculated lattice parameters and the details of electronic band structure are in good agreement with literature data. [43, 44, 45] For comparison, the lattice parameters calculated using ADF/BAND are **a** = 1.615 Å and **b** = 2.871 Å, the details of electronic band structure are almost the same as in Fig. (d). Therefore, both methods yield consistent results. In the following, we will mainly discuss the results obtained using QUANTUM ESPRESSO unless otherwise stated.

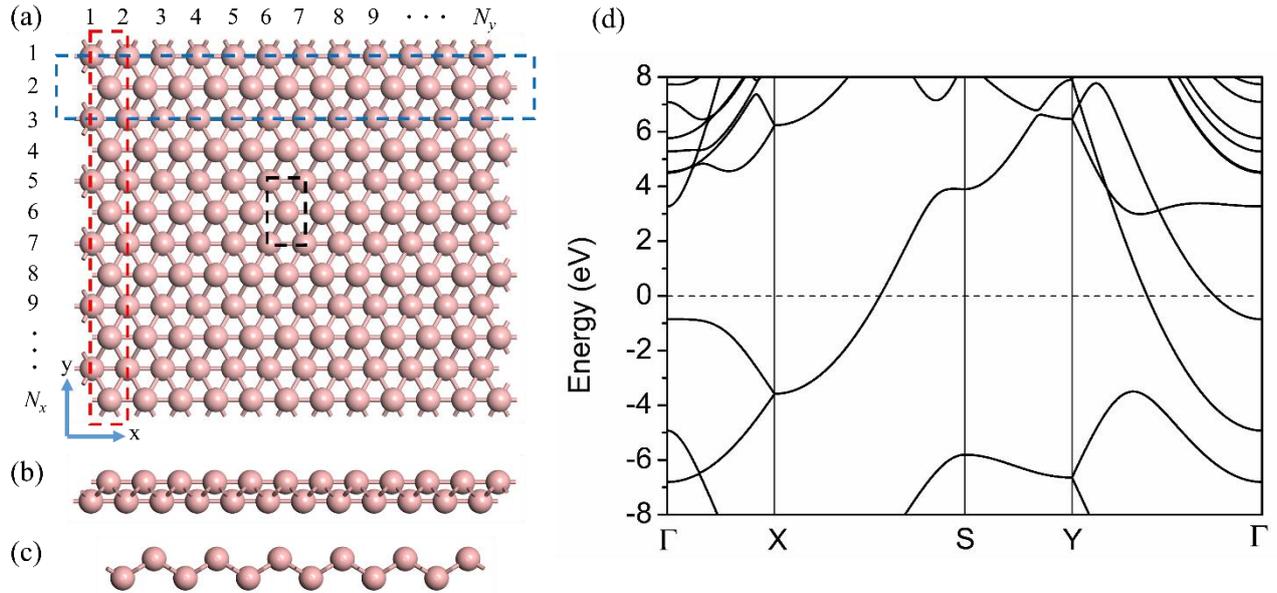

Figure 1. (a) Top view (b) side view along y-axis and (c) side view along x-axis of borophene, (d) electronic band structure of borophene with Fermi level shifted to 0 eV. The pink balls denote boron atoms. BxNRs and ByNRs are represented by the red and blue dashed rectangular, respectively. The unit cell of borophene used in the calculation is denoted by the black dashed rectangular.

As shown in Fig. 1(a), two varieties of BNR are obtained by cutting borophene along x-axis or y-axis, namely, BxNR and ByNR, respectively. Because of the peculiar geometric structure of borophene, cutting along either x-axis or y-axis results in only zigzag edge configuration. This is in contrast to graphene in which both zigzag and armchair edge configurations are possible depending on the cutting direction. Such BNRs are hereafter denoted as $N_x$-BxNR and $N_y$-ByNR with $N_x$ and $N_y$ characterizing the ribbon width in the units of respective lattice constant

(Fig. 1 (a)). In this work, $N_x$ and $N_y$ vary from 4 to 15, therefore the maximum ribbon width will be greater than 4 nm, which is experimentally attainable. In the following calculations, the lattice parameter for BxNRs and ByNRs are fixed at 1.616 Å and 2.868 Å, respectively.

**3.2 Geometric structure and stability of pristine BxNRs and ByNRs.**
Similar to the calculations of GNRs [26], three different *edge spin configuration*, namely, FM, AFM and NM were used as the initial states to verify the geometric structure and the stability of BNRs. Since the geometric structure does not vary much with the ribbon width, 8-BxNR and 8-ByNR are hereafter chosen as the representative. Figure 2 (a)-(d) depict the equilibrium structures of 8-BxNR and 8-ByNR. Spin polarized calculations show that BxNRs are NM, while ByNRs are nearly degenerate for AFM and FM state from the geometric point of view. For 8-ByNR, the bond lengths between B1 and B2 (B3) atoms are 1.631 Å (1.641 Å) for AFM state and 1.629 Å (1.642 Å) for FM state, respectively. In NM 8-ByNR, the corresponding B1-B2 and B1-B3 bond lengths are 1.578 Å and 1.657 Å, which is slightly different from the AFM and FM states.

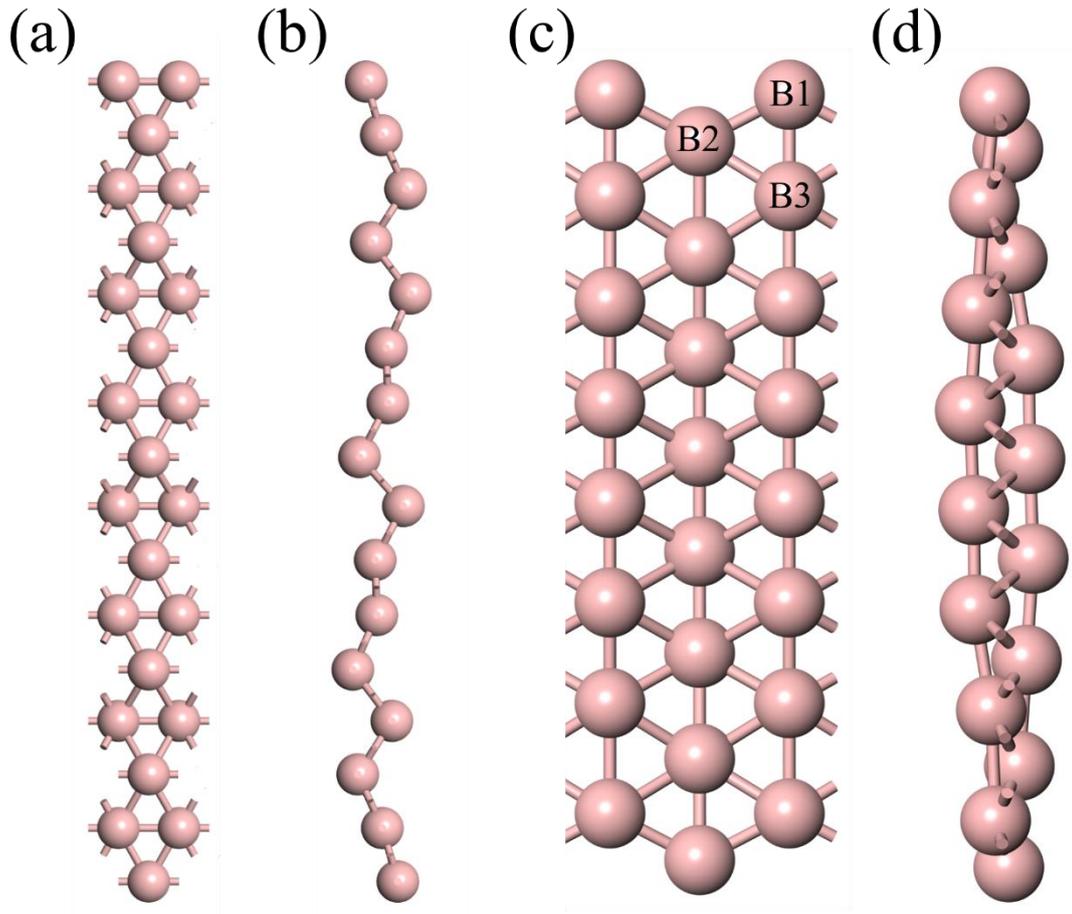

Figure 2. (a) Top view and (b) side view along x-axis of 8-BxNR, (c) top view and (d) side view along y-axis of 8-ByNR. The pink balls denote boron atoms.

To quantitatively verify the stability of BNRs, average binding energy (ABE) was calculated. ABE of the BNRs is defined as:

$$\text{ABE} = (n \times E_B - E_{\text{Total}}) / n \qquad (1)$$

where $E_{\text{Total}}$ is the total energy of the system, $E_B$ the energy of a single boron atom, and n the total number of boron atom. The calculated ABE are presented in Fig. 3. As shown, the values of ABE are positive for both

BxNRs and ByNRs, indicating these nanoribbon varieties are energetically stable. With the increase of ribbon width, the ABE is getting closer to the value of the two dimensional borophene (5.73 eV). With $N_y$ going from 4 to 15, the ABE of ByNRs increases from 5.30 eV to 5.60 eV, closer to the borophene value (5.73 eV); while the ABE of BxNRs remains nearly 5.65 eV. As can be seen from Fig. 3, the ground state of ByNRs is magnetic, either AFM of FM.

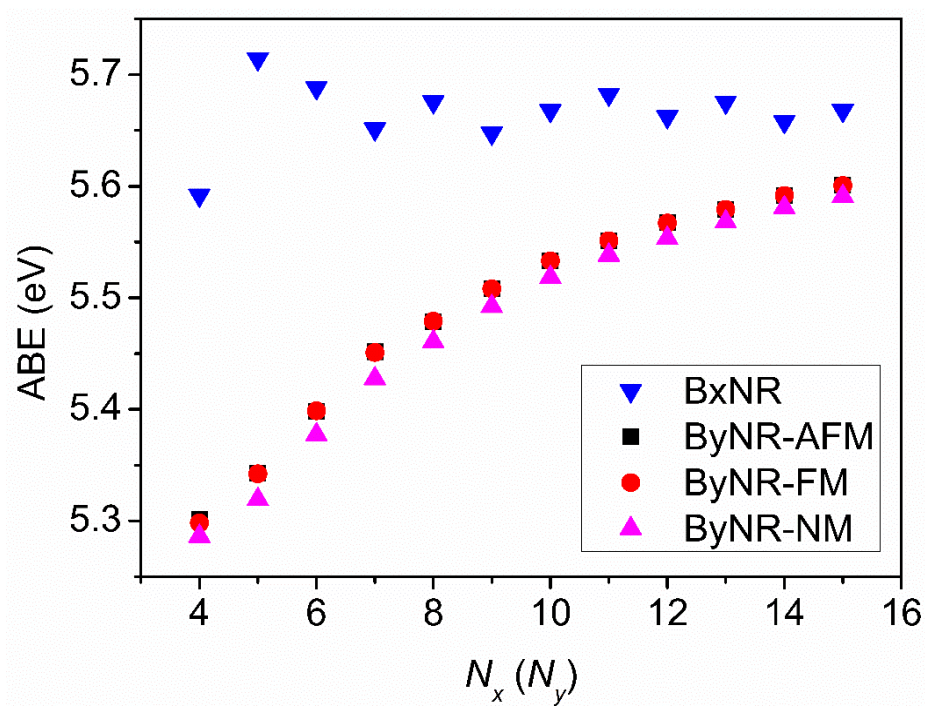

Figure 3. Average binding energy of pristine BNRs.

In order to determine which edge spin configuration is the ground state for ByNRs, the energy difference between AFM and FM states of ByNRs as a function of ribbon width $N_y$ is plotted in Fig. 4. With increasing $N_y$, the energy difference between AFM and FM state decreases rapidly. As

*Ny* increases from 5 to 10, the ground state is alternating between AFM and FM state. This interesting phenomenon, once experimentally verified, may have important technique implication. When *Ny* is larger than 10, tthe energy difference is less than 1 meV, indicating the AFM and FM states tend to be equally stable at a larger ribbon width, a reminiscent of ZGNRs. [26, 60] Also shown in Fig. 4 is the energy difference between the magnetic ground state and the NM state. As shown, the magnetic ground state is much lower in energy than the NM state, agreeing well with our ABE results.

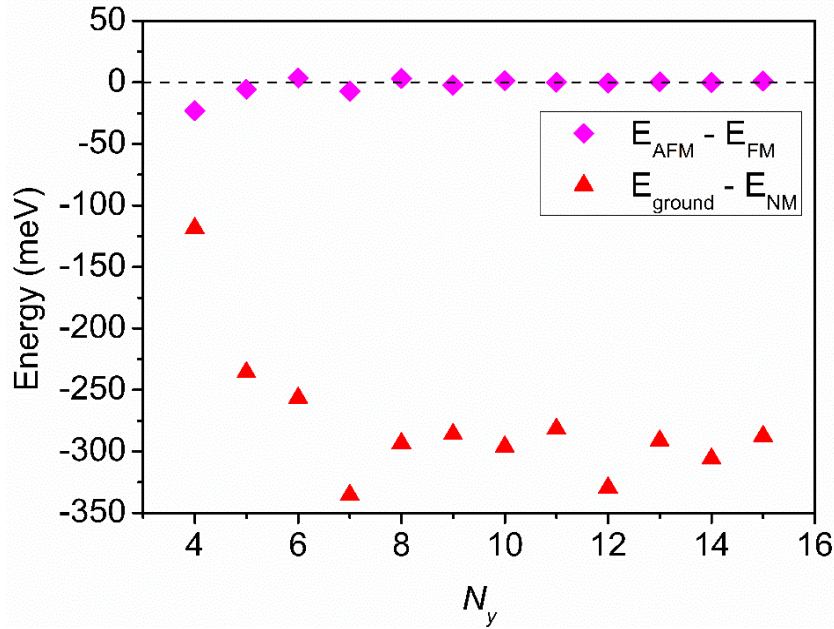

Figure 4. Energy difference between AFM and FM state and that between ground state and the NM state of pristine ByNRs.

**3.3 Electronic structure of pristine BxNRs and ByNRs.** The electronic

band structures of 8-BxNR and 8-ByNR are shown in Fig. 5 (a)-(d), respectively. As shown, 8-BxNR and 8-ByNR are metallic because of the band crossing at the Fermi level. In the case of AFM 8-ByNR, both spin up and spin down electrons have the same energy eigenvalues, indicating zero magnetization. It is interesting to compare Fig. 1 (d) and Fig. 5(c), the most notable difference is the band gap in borophene from Γ to X point in k-space is gone in 8-ByNR. An immediate question arises as to whether an external field, say, strain, can restore the band gap in ByNR. In the case of FM 8-ByNR, the energy difference between spin up and spin down electrons at the Fermi level is substantial, suggesting spontaneous magnetization.

Interestingly, we found that the electronic band structure of ByNRs exhibits a "re-entry" behavior with the ribbon width ($N_y$). Both AFM and FM 7-ByNR exhibit a semiconductor band structure (Fig. 5 (f)-(g)) whereas all other BNR varieties remain metallic. The AFM 7-ByNR has an indirect band gap of 0.26 eV (Fig. 5 (f)). In FM 7-ByNR, an indirect band gap is also observed with a band gap of 0.20 eV (Fig. 5 (g)). The band gaps are similar to the values obtained using ADF/BAND, which are 0.28 eV and 0.24 eV for AFM 7-ByNR and FM 7-ByNR, respectively. In BxNRs, the electrical conductance systematically increases with an increase of $N_x$ as there are more bands (conduction channels) crossing the

Fermi level.

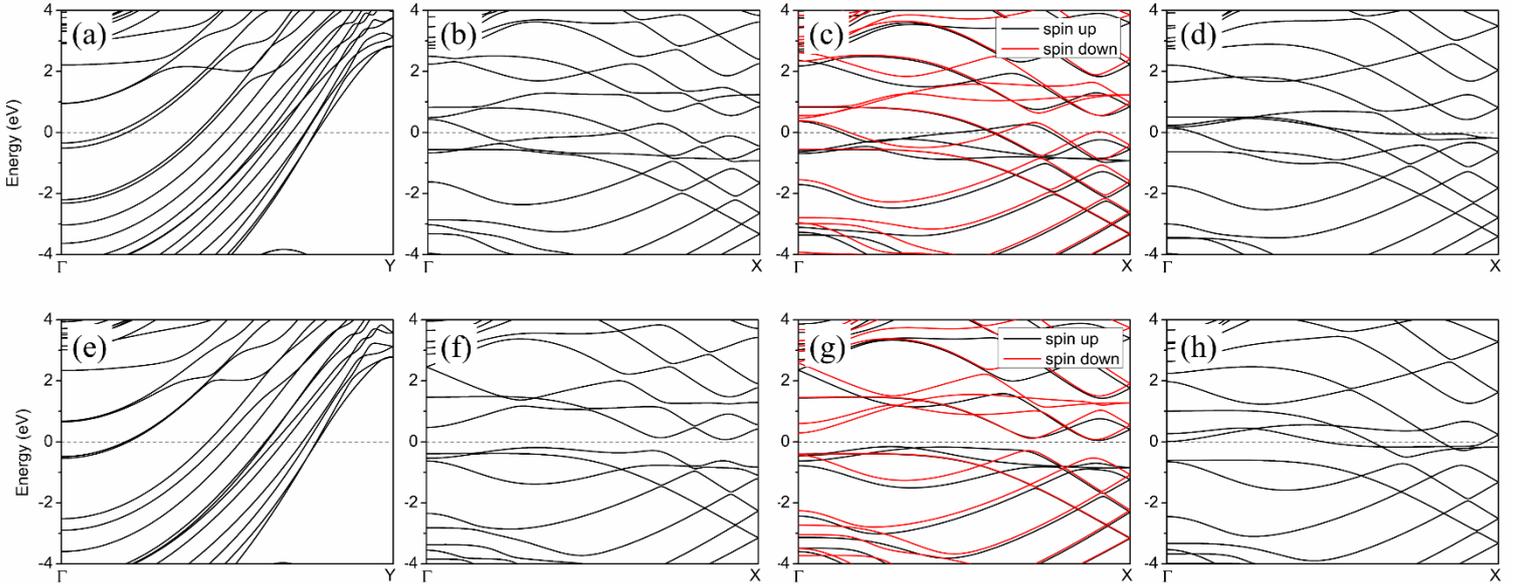

Figure 5. The band structure of (a) 8-BxNR, (b) AFM state of 8-ByNR, (c) FM state of 8-ByNR, (d) NM state of 8-ByNR, (e) 7-BxNR, (f) AFM state of 7-ByNR, (g) FM state of 8-ByNR, (h) NM state of 8-ByNR.

**3.4 Magnetic properties of pristine BxNRs and ByNRs.** In ByNRs, spin polarization calculations suggest that there may be spontaneous magnetization. Figure 6 (a) and (b) depict the spin density distribution of AFM and FM 8-ByNR obtained using ADF/BAND, respectively. As seen, finite magnetic moment mainly distributes at edge atoms while the magnetic moments are very small in the interior of BNRs. In the AFM state, spin configuration between two edges are AFM-like, i.e., the two edges have opposite spin orientation, leading the total magnetization to be zero. In contrast, in the FM state, the edge spin configuration is FM-like, resulting in a non-zero magnetization. Figure 6 (c) shows the variation of the total magnetic moment in FM 8-ByNR as a function of the ribbon width $N_y$. The total magnetic moment increases from 1.14 $\mu_B$ ($N_y = 4$) to 2.01 $\mu_B$ ($N_y = 15$), peaking at 2.06 $\mu_B$ ($N_y = 13$). The general trend of total magnetic moment as a function of ribbon width is almost the same as calculated by ADF/BAND, a peak also occurs at $N_y = 13$.

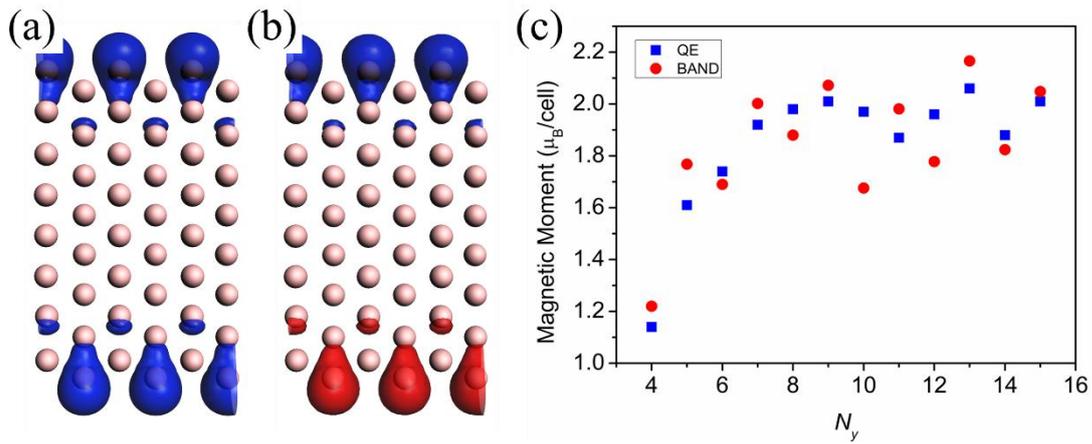

Figure 6. (a) Total magnetic moment of FM ByNR as a function of ribbon width $N_y$, and spin density distribution of (b) AFM 8-ByNR and (c) FM 8-ByNR. The pink balls denote boron atoms. The blue and red represent spin up and spin down density, respectively.

Apparently, the spontaneous magnetization of ByNRs is closely related to the edge states. As shown in Fig. 6 (a) and (b), the local magnetic moment is large on the edge atoms while it decreases obviously in the interior of ByNR. The profile of the local spin density strongly suggests that the dangling bonds on edges are from the *p*-electrons of B atoms.

To verify if the spontaneous magnetization of ByNR is indeed caused by the edge states, it is necessary to study the hydrogenation effect on the BNRs. In hydrogenation, one hydrogen atom is added on each edge to saturate the dangling bonds. To distinguish from the pristine BNRs discussed above, the hydrogenated BNRs are hereafter labeled as hBxNR or hByNR. In the remainder of this article, we will study the geometric structure, stability, electronic and magnetic properties of hBxNR (hByNR) for $N_x$ ($N_y$) = 4 to 15.

**3.5 Hydrogenation effects on the geometric structure and stability of BxNR and ByNR.** The fully optimized structures of 8-hBxNR and 8-hByNR are shown in Fig. 7. As for the hBxNRs, the changes mainly occur on the edge boron atoms: the edge atoms of pristine BxNR tend to be in the same plane whereas the edge atoms of hBxNR form a corrugated quasi-planar structure. Nonetheless, upon hydrogenation, the bond length between B1 and B2 atom increased from 1.629 Å to 1.690 Å in 8-hByNRs and the bond length between B1 and B3 atoms slightly decreased from 1.642 Å to 1.636 Å accordingly.

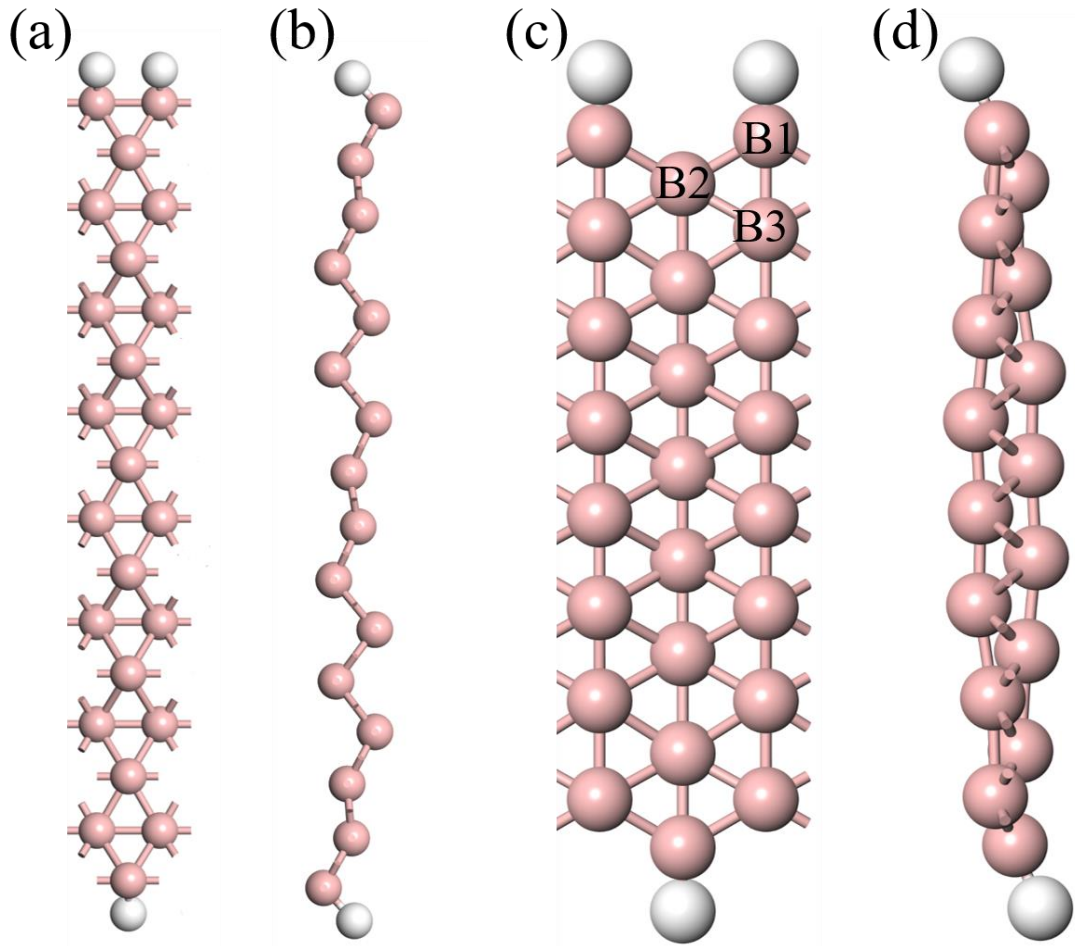

Figure 7. (a) Top view and (b) side view along x-axis of optimized geometric structure of 8-hBxNR, (c) top view and (d) side view along y-axis of optimized geometric structure of 8-hByNR. The pink and white balls denote boron and hydrogen atoms, respectively.

In order to examine the hydrogenation effects on the stability of BxNRs and ByNRs, ABE is again employed to study the stability of the system while we employ binding energy (BE) to examine if the process BNRs + $H_2$ = hBNRs is energetically favorable.

$$\text{ABE} = (n \times E_B + m \times E_H - E_{total}) / (n + m) \qquad (2)$$

where $E_{total}$ is the total energy of the system, $E_B$ is the energy of a single boron atom, $E_H$ is the energy of a single hydrogen atom, and n and m are the total number of boron atom and hydrogen atom, respectively.

$$BE = E_{BNR} + E_{H_2} - E_{total} \qquad (3)$$

where $E_{total}$ is the total energy of the system, $E_{BNR}$ the energy of pristine BNR and $E_{H_2}$ the energy of a hydrogen molecule.

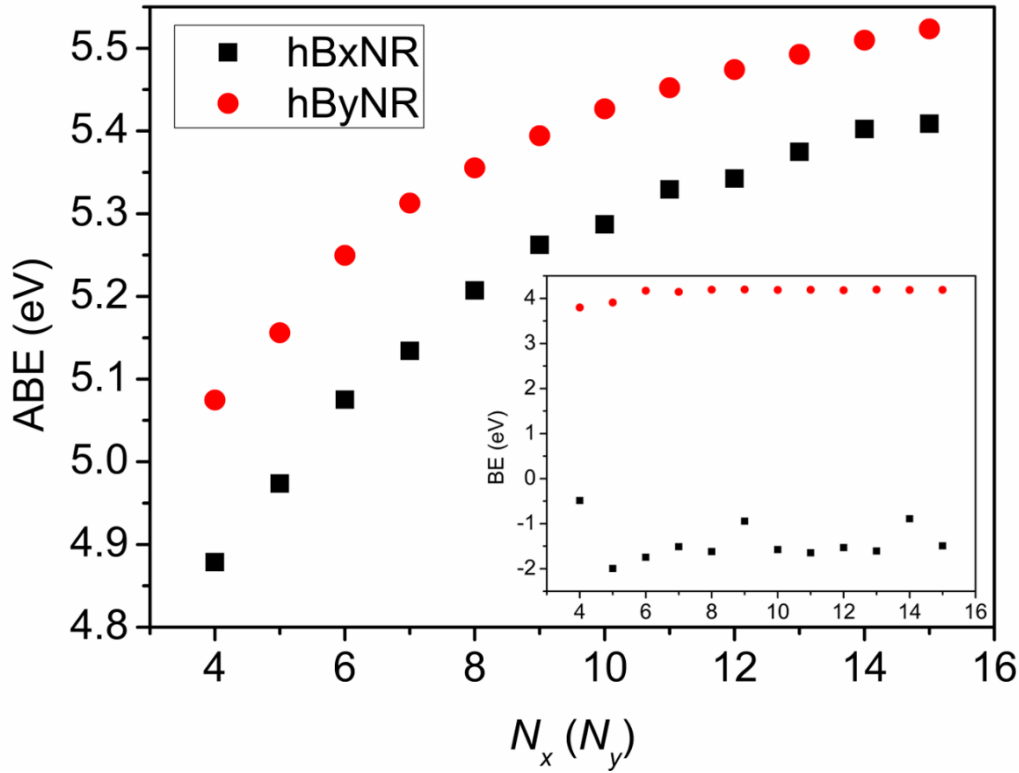

Figure 8. Average binding energy of hBNRs as a function of ribbon width $N_x$ ($N_y$). Insert: Binding energy of hBNRs as a function of ribbon width $N_x$ ($N_y$).

As shown in Fig. 8, though the ABE slightly decreases compared with the pristine BNRs, the ABE values of hBNRs are still positive, implying the structures are energetically favorable. In addition, the ABE of hBNRs shows the same trend as in the pristine BNRs (Fig. 3 and Fig. 8), in which the value of ABE increases with increasing ribbon width. The positive BE values indicate adding a hydrogen molecule to saturate the dangling bond is energetically favorable for hByNRs. In contrast, adding a hydrogen molecule to BxNRs is energetically unfavorable. This could result from the edge boron atoms in the pristine BxNRs are already saturated, which could help explain why pristine BxNRs are NM.

**3.6 Hydrogenation effects on the electronic structures and magnetic properties of BxNRs and ByNRs.**

The electronic band structure of the 8-hBxNR and 8-hByNR are shown in Fig. 9 (a) and (b). The major feature of the band structure, namely the metallicity, is preserved upon hydrogenation. In the pristine ByNRs, it is found that the band structure varies significantly with the ribbon width, both AFM and FM 7-ByNR are semiconductor with an indirect band gap. For hByNRs, with increasing $N_y$ from 4 to 15, hByNR changes first from metal to semiconductor and then to metal again. The 7-hByNR is a semiconductor with an indirect band gap ~ 0.21 eV, which slightly decreases compared to the AFM 7-ByNR. In addition, compared with the c pristine BxNRs, the conductance of hBxNRs increases with the ribbon width as there are more bands crossing the Fermi level.

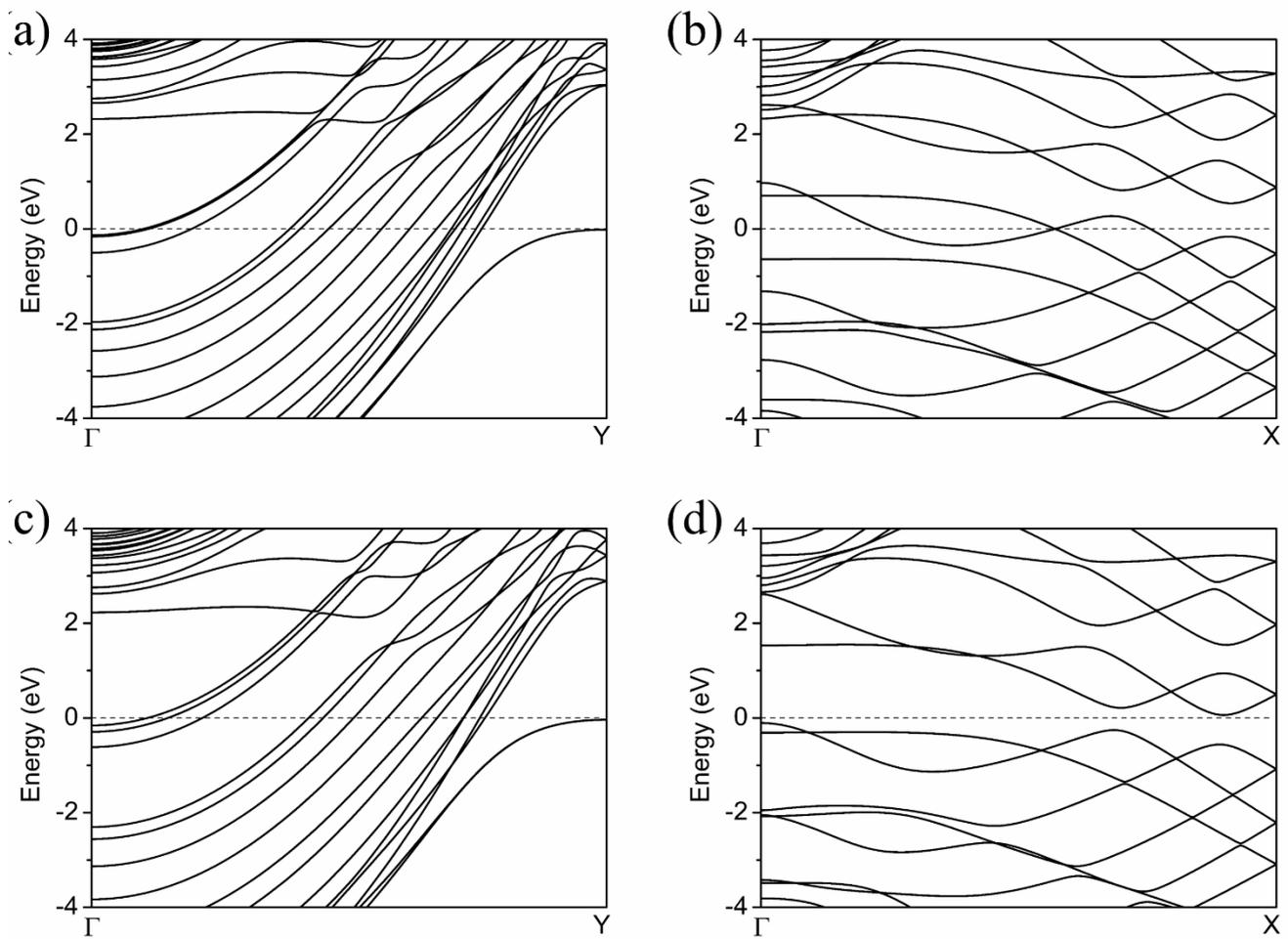

Figure 9. Calculated band structure of (a) 8-hByNR, (b) 8-hBxNR and (c) 7-hByNR.

Notably, upon hydrogenation, all BNRs adopt a NM ground state regardless the orientation and ribbon width. This thus confirms the key role of edge states in the magnetism of BNRs.

**Conclusion**

In analogy with GNRs, systematical study on the geometric, electronic, and magnetic properties of BNRs with orientation along x- and y-axis and the ribbon width $N_x$ ($N_y$) = 4 to 15 was carried out using DFT calculations. Spin polarized calculations demonstrated that BxNRs are NM while ByNRs adopt a magnetic ground state, either FM or NM state depending on the specific ribbon width. The edge states play a key role in the magnetic properties but less so in the electronic properties. Interestingly, pristine ByNR undergoes a metal-semiconductor-metal transition at $N_y$ = 7. Though BxNRs remain metallic regardless the $N_x$ value, the conductance of BxNRs systematically increases with increasing $N_x$. Hydrogenation turns the magnetic ground state to NM for all BNR varieties herein studied, illustrating the importance of the edge state on the magnetic properties. Upon hydrogenation, the metal-semiconductor-metal transition is preserved in hByNRs at $N_y$ = 7.

**Acknowledgements**

The authors would like to acknowledge the support of NSF DMR 1307740 and the discussion with Dr. Don Liebenberg.

**List of Figures**

**Figure 1.** (a) Top view (b) side view along y-axis and (c) side view along

x-axis of borophene, (d) electronic band structure of borophene with Fermi level shifted to 0 eV. The pink balls denote boron atoms. BxNRs and ByNRs are represented by the red and blue dashed rectangular, respectively. The unit cell of borophene used in the calculation is denoted by the black dashed rectangular.

**Figure 2.** (a) Top view and (b) side view along x-axis of 8-BxNR, (c) top view and (d) side view along y-axis of 8-ByNR. The pink balls denote boron atoms.

**Figure 3.** Average binding energy of pristine BNRs.

**Figure 4.** Energy difference between AFM and FM state and that between ground state and the NM state of pristine ByNRs.

**Figure 5.** The band structure of (a) 8-BxNR, (b) AFM state of 8-ByNR, (c) FM state of 8-ByNR, (d) NM state of 8-ByNR, (e) 7-BxNR, (f) AFM state of 7-ByNR, (g) FM state of 8-ByNR, (h) NM state of 8-ByNR.

**Figure 6.** (a) Total magnetic moment of FM ByNR as a function of ribbon width $N_y$, and spin density distribution of (b) AFM 8-ByNR and (c) FM 8-ByNR. The pink balls denote boron atoms. The blue and red represent spin up and spin down density, respectively.

**Figure 7.** (a) Top view and (b) side view along x-axis of optimized geometric structure of 8-hBxNR, (c) top view and (d) side view along y-axis of optimized geometric structure of 8-hByNR. The pink and white balls denote boron and hydrogen atoms, respectively.

**Figure 8.** Average binding energy of hBNRs as a function of ribbon width $N_x$ ($N_y$). Insert: Binding energy of hBNRs as a function of ribbon width $N_x$ ($N_y$).

**Figure 9.** Calculated band structure of (a) 8-hByNR, (b) 8-hBxNR and (c) 7-hByNR.